\documentclass[prd,aps]{revtex4}
\newcommand{\D}{\displaystyle}
\newcommand{\p}[4]{#1^{#3}_{#2,\,#4}}
\begin{document}
\input psfig
\title{Discrete space-time}

\author{Rodolfo Gambini$^1$, Jorge Pullin$^2$} \affiliation{1.
  Instituto de F\'{\i}sica, Facultad de Ciencias, Igu\'a esq. Mataojo,
  Montevideo, Uruguay\\
  2. Department of Physics and Astronomy, Louisiana State University,
  Baton Rouge, LA 70803 USA}

 \pacs{4.65} 

\begin{abstract}
  
  We review recent efforts to construct gravitational theories on
  discrete space-times, usually referred to as the ``consistent
  discretization'' approach. The resulting theories are free of
  constraints at the canonical level and therefore allow to tackle
  many problems that cannot be currently addressed in continuum
  quantum gravity. In particular the theories imply a natural method
  for resolving the big bang (and other types) of singularities and
  predict a fundamental mechanism for decoherence of quantum states
  that might be relevant to the black hole information paradox.  At a
  classical level, the theories may provide an attractive new path for
  the exploration of issues in numerical relativity.  Finally, the
  theories can make direct contact with several kinematical results of
  continuum loop quantum gravity.  We review in broad terms several of
  these results and present in detail as an illustration the classical
  treatment with this technique of the simple yet conceptually
  challenging model of two oscillators with constant energy sum. 

  \end{abstract}
  \maketitle

\section{Introduction}

The idea that space-time might be discrete has arisen at various
levels in gravitational physics. On one hand, some approaches
hypothesize that at a fundamental level a discrete structure underlies
space-time.  Other approaches start with a continuum theory but upon
quantization discrete structures associated with space-time emerge.
Finally, discretizations are widely used in physics, and in gravity in
particular, as a calculational tool at two levels: a) at the time of
numerically computing predictions of the theory (classical and quantum
mechanically) and b) as a regularization tool for quantum
calculations.

Whatever the point of view that may lead to the consideration of a
discrete space-time, the formulation of gravitational theories on such
structures presents significant challenges. At the most immediate
level, the presence of discrete structures can conflict with
diffeomorphism invariance, a desirable property of gravitational
theories. This manifests itself in various ways. For instance, if one
simply proceeds to discretize the equations of motion of general
relativity, as is common in numerical relativity applications, one
finds that the resulting equations are inconsistent: the evolution
equations do not preserve the constraints, as they do in the
continuum. In another example, if one considers the discretization of
the constraints of canonical quantum gravity, the resulting discrete
constraints fail to close an algebra, which can be understood as
another manifestation of the inconsistency faced in numerical
relativity.

A new viewpoint has recently emerged towards the treatment of theories
on discrete space-times. At the most basic level, the viewpoint
advocates discretizing the action of the theory and working out the
resulting equations of motion rather than discretizing the equations
of motion directly. The resulting equations of motion stemming from
the discrete action are generically guaranteed to be consistent.  So
immediately the problem of consistency is solved. This has led to the
approach being called the ``consistent discretization'' approach.
This approach has been pursued in the past in numerical approaches
to unconstrained theories and is known as ``variational integrators''
(see \cite{Marsd} for a review) and it appears to have several 
desirable properties. Constrained systems have only been considered
if they are holonomic (i.e. they only depend on the configuration
variables), although see \cite{mclc} 
for some recent results for anholonomic
constraints.

In spite of these positive prospects, the resulting theories have
features that at first sight may appear undesirable. For instance,
quantities that in the usual continuum treatment are Lagrange
multipliers of first class constrained systems (and therefore freely
specifiable), become determined by the equations of motion.  The
equations that determine the Lagrange multipliers in general may have
undesirable complex solutions.

On the other hand, the approach has potentially very attractive
features: equations that in the continuum are constraints among the
dynamical variables become evolution equations that are automatically
solved by the scheme.  Having no constraints in the theory profoundly
simplifies things at the time of quantization. The conceptually hard
problems of canonical quantum gravity are almost entirely sidestepped
by this approach. For instance, one can introduce a relational
description in the theory and therefore solve the ``problem of time''
that created so much trouble in canonical quantum gravity. The
resulting relational description naturally implies a loss of unitarity
that may have implications for the black hole information puzzle. The
discrete theories also have a tendency to avoid singularities, since
the latter usually do not fall on the computational grid. At a quantum
level this implies that singularities have zero probability. This
provides a singularity avoidance mechanism that is distinct from the
one usually advocated in loop quantum cosmology.  From the point of
view of numerical relativity, the resulting evolution schemes preserve
the constraints of the continuum theory to a great degree of accuracy,
at least for solutions that approximate the continuum limit. This is
different from usual ``free evolution'' schemes which can converge to
continuum solutions that violate the constraints.  As we will see, we
are still somewhat away from being able to advocate that the resulting
schemes can be competitive in numerical relativity.  But given that
enforcing the constraints has been identified by some researchers as
the main obstacle to numerical relativity, the proposed 
schemes deserve some
consideration.

An aspect of presupposing a discrete structure for space-time that may
also appear undesirable is that in loop quantum gravity the discrete
structures only emerge after quantization. The initial formulation of
the theory is in the continuum. Therefore there is the risk that the
new viewpoint may not be able to make contact with the many attractive
kinematical results of loop quantum gravity. We will see that a
connection is in fact possible, and it retains the attractive aspects
of both approaches.

In this paper we would like to review the ``consistent
discretization'' approach to gravitational theories. In section II we
will outline the basics of the strategy. In section III we concretely
apply the method to a simple yet important model problem so the
reader can get a flavor of what to expect from the method classically.
In section IV we discuss cosmological models and in section V
the introduction of a relational time and the black hole information
puzzle. In section VI we discuss connections with continuum loop
quantum gravity. We end with a discussion.

\section{Consistent discretizations}

We start by considering a continuum theory representing a mechanical
system. Although our ultimate interest is in field theories, the
latter become mechanical systems when discretized. Its Lagrangian will
be denoted by ${\hat L(q^a,\dot{q}^a)}$, $a=1\ldots M$.  This setting
is general enough to accommodate, for instance, totally constrained
systems. In such case $\dot{q}$ will be the derivative of the
canonical variables with respect to the evolution parameter. It is
also general enough to include the systems that result from
formulating on a discrete space-time lattice a continuum field theory.

We discretize the evolution parameter in intervals (possibly varying
upon evolution) $t_{n+1}-t_n=\epsilon_n$ and we label the generalized
coordinates evaluated at $t_n$ as $q_n$. We define the discretized
Lagrangian as

\begin{equation} 
L(n,n+1) \equiv L(q^a_n,q^a_{n+1}) \equiv \epsilon_n{\hat
L}(q^a, \dot q^a)\label{disc} \end{equation}
where
\begin{equation} q^a=q^a_n \quad \hbox{and} \quad \dot{q}^a \equiv
{{q^a_{n+1}-q^a_n}\over \epsilon_n}. \end{equation}

Of course, one could have chosen to discretize things in a different
fashion, for instance using a different approximation for the
derivative, or by choosing to write the continuum Lagrangian in terms
of different variables. The resulting discrete theories generically
will be different and will approximate the continuum theory in
different ways. However, given a discrete theory, the treatment we
outline in this paper is unique.

The action can then be written as
\begin{equation}
S=\sum_{n=0}^N L(n,n+1).
\end{equation}

It should be noted that in previous treatments \cite{DiGaPu,GaPu} we
have written the Lagrangian in first order form, i.e. $L=\int dt
\left(p\dot{q}-H(p,q)\right)$. It should be emphasized that this is
contained as a particular case in the treatment we are presenting in
this paper. In this case one takes both $q$ and $p$ to be configuration
variables, and one is faced with a Lagrangian that involves $q_n,p_n$
and $q_{n+1}$ as variables, being independent of $p_{n+1}$.

If the continuum theory is invariant under
reparameterizations of the evolution parameter, one can show that the
information about the intervals $\epsilon_n$ may be absorbed in the
Lagrange multipliers. In the case of standard mechanical systems it is
simpler to use  an invariant interval $\epsilon_n=\epsilon$.  

The Lagrange equations of motion are obtained by requiring the
action to be stationary under variations of the configuration
variables $q^a$ fixed at the endpoints of the evolution
interval $n=0,n=N+1$,
\begin{equation}
{\partial L(n,n+1) \over \partial  q^a_{n}}+{\partial L(n-1,n)
\over \partial q^a_{n}}=0. \label{lagra}
\end{equation}

We introduce the following definition of the canonically conjugate
momenta of the configuration variables,
\begin{eqnarray}
 p^a_{n+1} &\equiv& {\partial L(n,n+1) \over \partial  q^a_{n+1}}
\label{1}\\
 p^a_{n} &\equiv& {\partial L(n-1,n) \over \partial  q^a_{n}}=-
{\partial L(n,n+1) \over \partial  q^a_{n}}\label{2}
\end{eqnarray}

Where we have used Eq. (\ref{lagra}). The equations (\ref{1}) and
(\ref{2}) define a canonical transformation for the variables
$q_n,p_n$ to $q_{n+1},p_{n+1}$ with a the type 1 generating function
$F_1= -L( q^a_n, q^a_{n+1})$. Notice that the evolution scheme is
implicit, one can use the bottom equation (since we are in the
non-singular case) to give an expression for $q_{n+1}$ in terms of
$q_n,p_n$, which in turn can be substituted in the top equation to get
an equation for $p_{n+1}$ purely in terms of $q_n,p_n$.

It should be noted that there are several other possible choices,
when going from the set of equations (\ref{1},\ref{2}) to 
an explicit evolution scheme (see \cite{jmp} for further details.)

The canonical transformation we introduced becomes singular as an
evolution scheme if $\D\left|{\partial^2 L(n,n+1) \over \partial
q^a_{n+1}\partial q^b_{n}}\right|$ vanishes. If the rank of the matrix
of second partial derivatives is $K$ the system will have $2(M-K)$
constraints of the form,
\begin{eqnarray}
\Phi_A( q^a_n, p^a_n)&=&0\\
\Psi_A( q^a_{n+1}, p^a_{n+1})&=&0.
\end{eqnarray}
And these constraints need to be enforced during evolution, which may
lead to new constraints. We refer the reader for the 
detailed Dirac analysis to \cite{jmp}. 

To clarify ideas, let us consider an example. 
The model consists of a parameterized free
particle in a two dimensional space-time under the influence of a
linear potential. The discrete Lagrangian is given by,
\begin{equation}
L_n\equiv L(q_n^a,\pi_n^a,N_n,q_{n+1}^a,\pi_{n+1}^a,N_{n+1}) =
\pi_n^a(q_{n+1}^a-q^a_n)-N_n[\pi_n^0+
\frac{1}{2}(\pi_n^1)^2+\alpha q_n^1] \label{la}.
\end{equation}
We have chosen a first order formulation for the particle. However,
this Lagrangian is of the type we considered in this paper, one 
simply needs to consider all variables, $q^a,\pi^a,N$ as configuration
variables. The system is clearly  singular since the $\pi's$ and
$N$ only appear at level $n$ (or in the continuum Lagrangian, their
time derivatives are absent). When considered as a Type I
generating function, the above Lagrangian leads to the equations
\begin{eqnarray}
\p{p}{\pi}{a}{n+1} &=& \frac{\partial L_n}{\partial \pi^a_{n+1}}
=0, \label{evol11}
\\
\p{p}{q}{a}{n+1} &=& \frac{\partial L_n}{\partial q^a_{n+1}}
=\pi_n^a, \label{evol12}
\\
\p{p}{N}{}{n+1} &=& \frac{\partial L_n}{\partial N_{n+1}} =0,
\label{evol3}
\end{eqnarray}
and
\begin{eqnarray}
\p{p}{\pi}{a}{n} &=& -\frac{\partial L_n}{\partial \pi^a_n}
=-(q_{n+1}^a-q_n^a)+ \pi_n^1 N_n \delta^a_1+ N_n \delta^a_0,
\label{evo21}
\\
\p{p}{q}{a}{n} &=& -\frac{\partial L_n}{\partial q^a_n}
=\pi_n^a+\delta^a_1\alpha N_n, \label{evo22}
\\
\p{p}{N}{}{n} &=& -\frac{\partial L_n}{\partial N_n}
=\pi_n^0+\frac{1}{2}(\pi_n^1)^2+\alpha q_n^1 \label{evo23}.
\end{eqnarray}
The constraints (\ref{evol11},\ref{evol3},\ref{evo22},\ref{evo23}) can
be imposed strongly to eliminate the $\pi$'s and the $N$'s and obtain
an explicit evolution scheme for the $q$'s and the $p_q$'s,
\begin{eqnarray}
q^0_n &=&   q^0_{n+1} - \frac{C_{n+1}}{\alpha
\p{p}{q}{1}{n+1}},
\\
q^1_n &=&  q^1_{n+1} - \frac{C_{n+1}}{\alpha},
\\
\p{p}{q}{0}{n} &=& \p{p}{q}{0}{n+1},
\\
\p{p}{q}{1}{n}&=& \p{p}{q}{1}{n+1} 
+ \frac{C_{n+1}}{\p{p}{q}{1}{n+1}},
\end{eqnarray}
and the Lagrange multipliers get determined to be,
\begin{equation}
  N_n = \frac{C_{n+1}}{\alpha \p{p}{q}{1}{n+1}},
\end{equation}
where $C_{n+1}=\p{p}{q}{0}{n+1} + (\p{p}{q}{1}{n+1})^2/2 +\alpha
q^1_{n+1}$. The evolution scheme runs backwards, one can construct a
scheme that runs forward by solving for $N$ and $\pi$ at instant $n$
when imposing the constraints strongly.
The two
methods yield evolution schemes of different functional form
since one propagates ``forward'' in time and the other ``backward''.
The inequivalence in the functional form stems from the fact that
the discretization of the time derivatives chosen in the Lagrangian
is not centered. It should be emphasized that if one starts from
given initial data and propagates forwards with the first system of
equations and then backwards using the second, one will return to
the same initial data.

So we see in the example how the mechanism works. It yields evolution
equations that usually are implicit as evolution schemes. The
equations are consistent. The Lagrange multipliers get determined by
the scheme and there are no constraints left on the canonical
variables. The evolution is implemented by a (non-singular) canonical
transformation. The number of degrees of freedom is larger than those
in the continuum. There will exist different sets of initial data that
lead to different solutions for the discrete theory but nevertheless
will just correspond to different discrete approximations and
parameterizations of a single evolution of the continuum theory.

\section{The Rovelli model}

To analyze the method in a simple ---yet challenging--- model we
consider the model analyzed by Rovelli \cite{Rovelli} in the context
of the problem of time in canonical quantum gravity: two harmonic
oscillators with constant energy sum. The intention of this section is
to illustrate how the method works and some of the expectations one
can hold when applying the method to more complex situations. The
model itself can obviously be treated with more straightforward
discretization techniques given its simplicity. The fact that the
method works well for the model should not be construed as proof that
it will be successful in other numerical applications. Current efforts
suggest that the technique is successfully applicable to Gowdy
cosmologies.

The model has canonical coordinates
$q^1,q^2,p^1,p^2$ with the standard Poisson brackets and a constraint
given by,
\begin{equation}
  C=\frac{1}{2} \left((p^1)^2+(p^2)^2+(q^1)^2+(q^2)^2\right)-M=0,
\end{equation}
with $M$ a constant. No Hamiltonian system can correspond to this
dynamical system since the presymplectic space is compact and
therefore cannot contain any $S\times R$ structure. Nevertheless, we
will see that the consistent discretization approach does yield
sensible results. This helps dispel certain myths about the consistent
discretization scheme. Since it determines Lagrange multipliers a lot
of people tend to associate the scheme as some sort of ``gauge
fixing''.  For this model however, a gauge fixing solution would be
unsatisfactory, since it would only cover a portion of phase space. We
will see this is not the case in the consistent discretization
scheme. We will also see that the evolution scheme is useful
numerically in practice.

We start by writing a discrete Lagrangian for the model,
\begin{equation}
L(n,n+1) = 
p^1_n \left( q^1_{n+1}-q^1_n\right)+
p^2_n \left( q^2_{n+1}-q^2_n\right)-
\frac{N_n}{2}\left(
(p^1_n)^2+(p^2_n)^2+(q^1_n)^2+(q^2_n)^2-2 M\right),
\end{equation}
and working out the canonical momenta for all the variables, i.e.,
$P_q^1,P_q^2,P_p^1,P_p^2$. One then eliminates the $p^{1,2}$
and the $P_p^{1,2}$ and is left with evolution equations for the
canonical pairs,
\begin{eqnarray}
  q^1_{n+1} &=& q^1_n+N_n \left(P^1_{q,n}-2 q^1_n\right)\\
  q^2_{n+1} &=& q^2_n+N_n \left(P^2_{q,n}-2 q^2_n\right)\\
  P^1_{q,n+1} &=&  P^1_{q,n} -N_n q^1_n\\
  P^2_{q,n+1} &=&  P^2_{q,n} -N_n q^2_n.
\end{eqnarray}

The Lagrange multiplier gets determined by the solution(s) of a 
quadratic equation, 
\begin{equation}
\left((q^1_n)^2+(q^2_n)^2\right)(N_n)^2-2\left(P^1_{q,n} q^1_n+
P^2_{q,n} q^2_n\right)N_n 
+\left(  P^1_{q,n}\right)^2+\left(  P^2_{q,n}\right)^2
+\left(q^1_n\right)^2+\left(q^2_n\right)^2-2M=0.
\label{quadratic}
\end{equation}

We would like to use this evolution scheme to follow numerically the
trajectory of the system. For this, we need to give initial
data. Notice that if one gives initial data that satisfy the
constraint identically at level $n$, the quadratic equation for the
lapse has a vanishing independent term and therefore the solution is
that the lapse $N$ vanishes (the nonvanishing root will be large and
far away from the continuum generically). To construct initial data
one therefore considers a set for which the constraint vanishes and
introduces a small perturbation on one (or more) of the
variables. Then one will have evolution. Notice that one can make the
perturbation as small as desired. The smaller the perturbation, the
smaller the lapse and the closer the solution will be to the
continuum.

For concreteness, we choose the following initial values for the
variables, $M=2$,
\begin{eqnarray}
  q^1_0 &=&0,\\
  q^2_0 &=&(\sqrt{3}-\Delta) \sin({\pi\over 4}),\\
  P^1_{q,0} &=& 1,\\
  P^1_{q,0} &=& (\sqrt{3}-\Delta) \cos({\pi\over 4}),
\end{eqnarray}
We choose the parameter $\Delta$ to be the perturbation, i.e.,
$\Delta=0$ corresponds to an exact solution of the constraint, for
which the observable $A=1/2$ (see below for its definition). The
evolution scheme can easily be implemented using a computer algebra
program like Maple or Mathematica.

Before we show results of the evolution, we need to discuss in some
detail how the method determines the lapse. As we mentioned, it is
obtained by solving the quadratic equation (\ref{quadratic}). This
implies that generically there will be two possible solutions and in
some situations they could be negative or complex. One can choose any
of the two solutions at each point during the evolution. It is natural
numerically to choose one ``branch'' of the solution and keep with
it. However, if one encounters that the roots become complex, we have
observed that it is possible to backtrack to he previous point in the
iteration, choose the alternate root to the one that had been used up
to that point and continue with the evolution. Similar procedure could
be followed when the lapse becomes negative. It should be noted that
negative lapses are not a problem per se, it is just that the
evolution will be retraced backwards. We have not attempted to correct
such retracings, i.e. in the evolutions shown we have only ``switched
branches'' whenever the lapse becomes complex. This occurs when the
discriminant in the quadratic equation (\ref{quadratic}) vanishes.

\begin{figure}[htbp]
  \centerline{\psfig{file=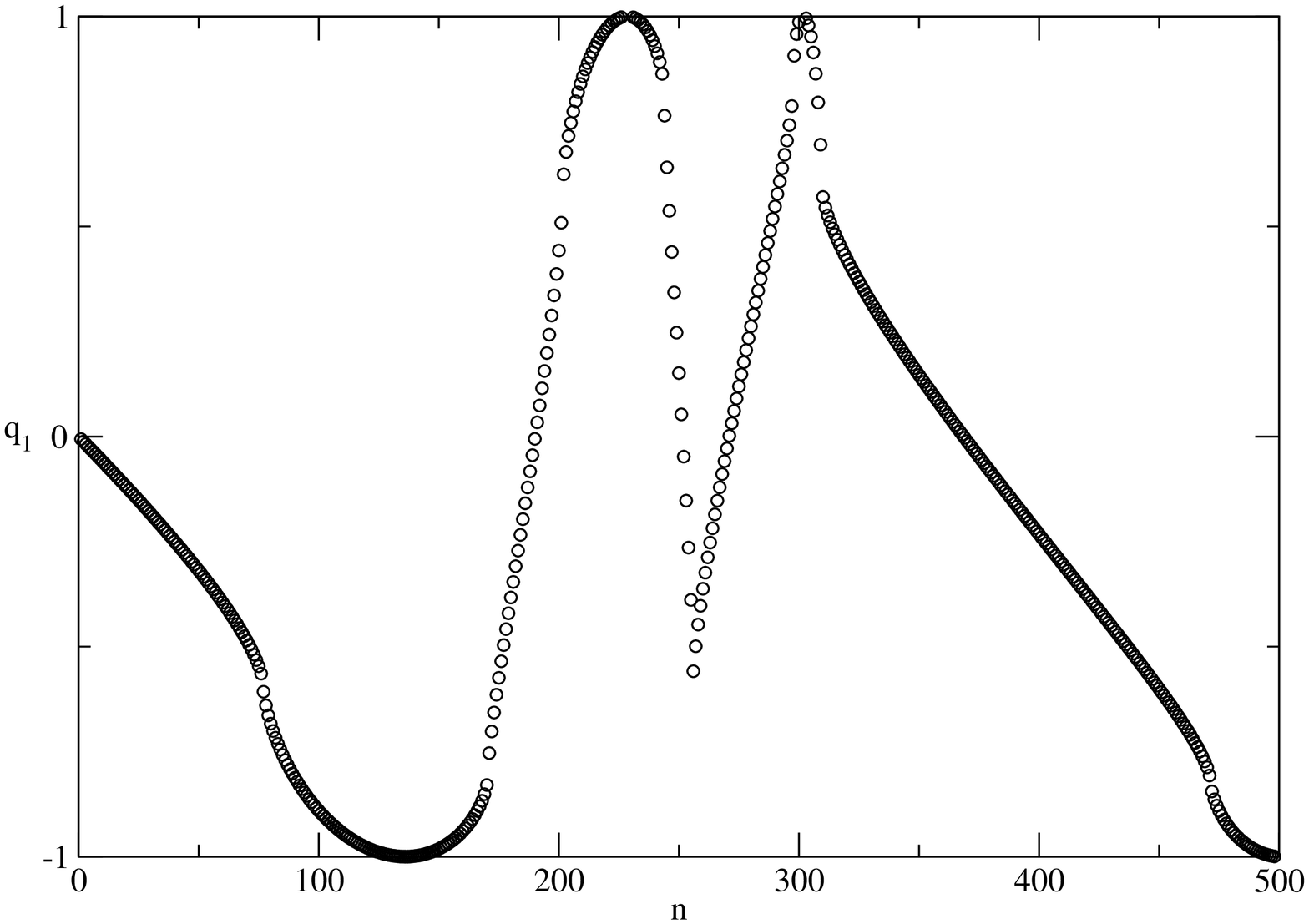,height=6cm}} \caption{The evolution
    of one of the variables of the oscillator model as a function of
    the discretization parameter $n$. The irregular nature of the
    curve is due to the fact that the lapse is a dynamical variable
    and therefore the rate of advance changes during evolution. The
    sharp feature around $n=250$ is due to the fact that the lapse
    becomes negative and the evolution runs backwards for a while,
    until about $n=310$.}  \label{figure1} \end{figure}

Figure (\ref{figure1}) shows the evolution of $q_1$ as a function of
$n$. The figure looks choppy since the ``rate of advance'' (magnitude
of the lapse) varies  during the evolution. 

Figure (\ref{figure1}) also exhibits that in a reparameterization
invariant theory like this one it is not too useful to plot
parameterization dependent quantities. One would have to exhibit
relational quantities that are true observables of the theory to
obtain physically relevant information. In this particular model,
Rovelli has given an explicit expression for a relational
(``evolving'') observable
\begin{equation}
  q^2(q^1) = \sqrt{M/A -1}\left[q^1 \cos\phi\pm \sqrt{2A-(q^1)^2}\sin\phi\right],
\end{equation}
where $A$ and $\phi$ are constants of the motion (``perennials''),
whose expression in terms of the coordinates is,
\begin{eqnarray}
  4A&=&2M+(p^1)^2-(p^2)^2+(q^1)^2-(q^2)^2,
  \tan\phi ={p^1 q^2-p^2 q^1\over p^1 p^2+q^2 q^1}.
\end{eqnarray}

\begin{figure}[htbp]
  \centerline{\psfig{file=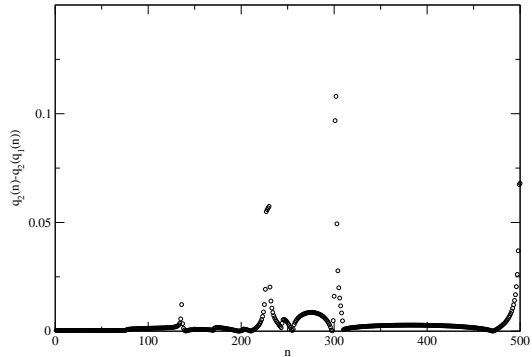,height=6.2cm}} \caption{The
    error in the evaluation of the relational observable in the
    discretized evolution, compared with respect to the exact
    continuum expression. We show the absolute error, but since the
    quantities are of order one it can also be understood as the
    relative error. The peaks in the error are due to the functional
    form of the observable involving the square root of $2A-q_1^2$.
    This vanishes when $q_1=\pm1$ (see previous plot) and this
    magnifies the error in the observable.}
  \label{figure2} \end{figure}

The relational observable gives an idea of the trajectory in
configuration space in a manner that is invariant under
reparameterizations.  In figure (\ref{figure2}) we show the error in
the evaluation of the relational observable with respect to the exact
expression of the continuum in our model.

This model has two independent ``perennials'' that can be
used to construct relational observables like the one we just
discussed. The first one of these perennials happens to be an
exact conserved quantity of the discretized theory. The relation
between perennials in the continuum and conserved quantities of the
discrete theory was further discussed in \cite{cosmo}. The perennial
in question is,
\begin{equation}
  O_1=p^1 q^2-p^2 q^1.
\end{equation}

\begin{figure}[htbp]
  \centerline{\psfig{file=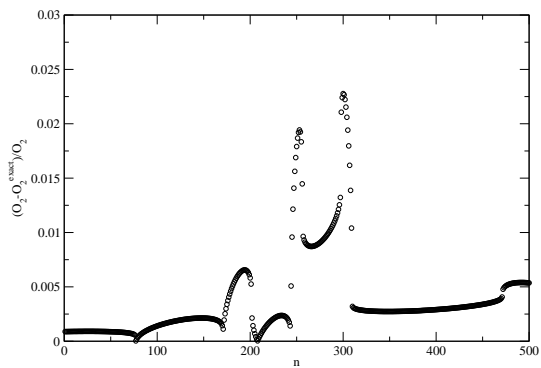,height=6.2cm}}
    \caption{The model has two ``perennials". One of them is an exact
    conserved quantity of the discrete theory, so we do not present a
    plot for it. The second perennial $(O_2)$ is approximately
    conserved. The figure shows the relative error in its computation
    in the discrete theory. It is worthwhile noticing that, unlike
what is usual in free evolution schemes, errors do not accumulate,
    they may grow for a while but later they might diminish. }
  \label{figure4}
\end{figure}

Another perennial is given by
\begin{equation}
  O_2=(p^1)^2-(p^2)^2+(q^1)^2-(q^2)^2.
\end{equation}
This quantity is not an exact conserved quantity of the discrete
model, it is conserved approximately, as we can see in figure
(\ref{figure4}).  We see that the discrete theory conserves the
perennial quite well in relative error and even though in intermediate
steps of the evolution the error grows, it decreases later.

\begin{figure}[htbp]
    \centerline{\psfig{file=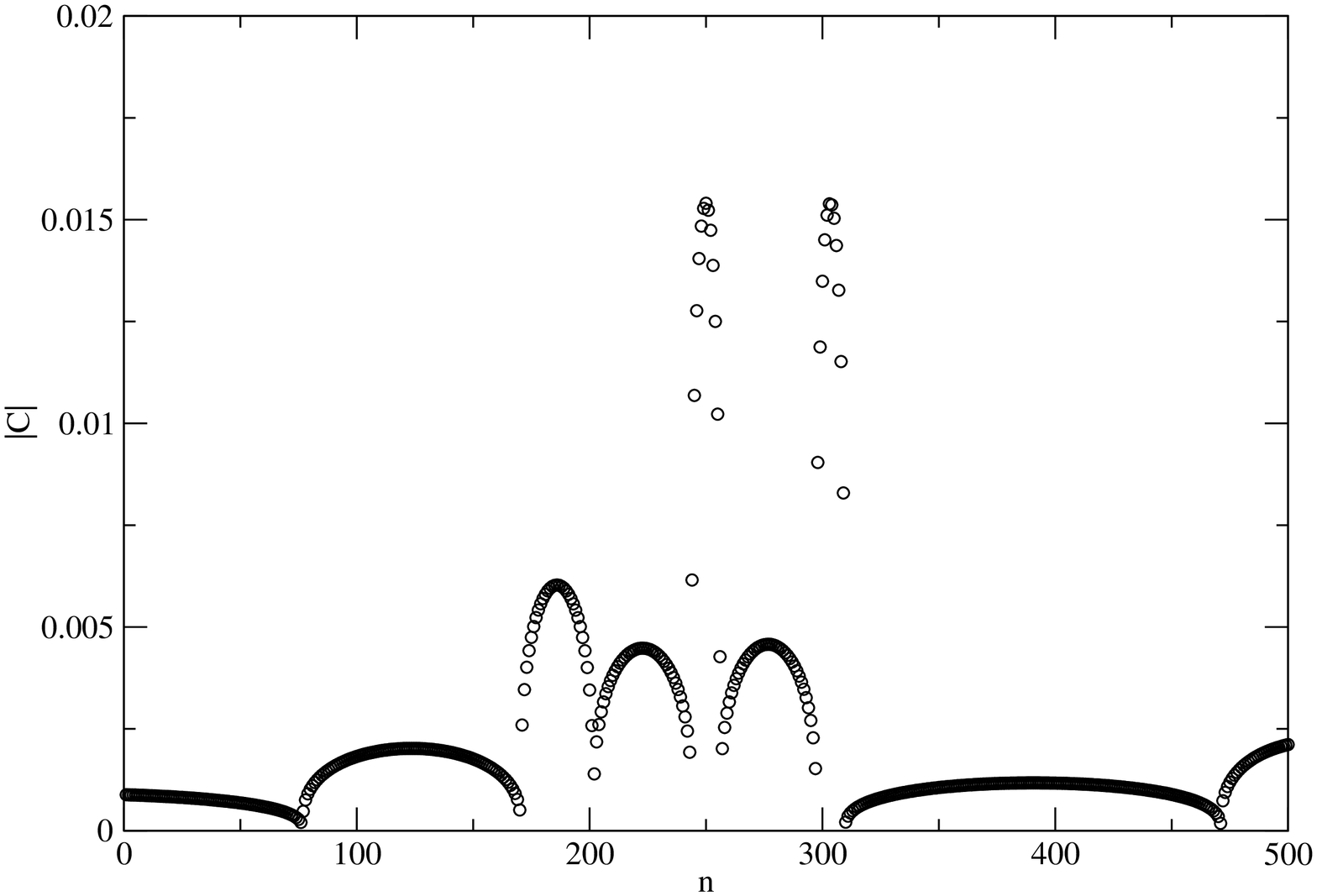,height=6.2cm}}
  \caption{Absolute value of the constraint of the continuum theory, evaluated in the discrete theory. The plot shows that the constraint of the continuum theory does not increase in the discrete theory as a function of evolution (a major desired goal in numerical relativity). The value of the constraint is also a measure of the error in the quantities of the discrete theory with respect to those of the continuum one (compare with the error in the observable in    figure 3, for example) that can be used to independently assess    the accuracy of the theory in convergence studies.}
  \label{figure5}
\end{figure}

In figure (\ref{figure5}) we depict the absolute value of the
constraint of the continuum theory as a function of discrete time
$n$. It is interesting to observe that in the discrete theory the
variables approximate the ones of the continuum with an error that is
proportional to the value of the constraint. Therefore the value of
the constraint can be taken as an indicator of how accurately one is
mirroring the continuum theory. It is a nice feature to have such an
error indicator that is independent of the knowledge of the exact
solution. This is clearly exhibited by contrasting (\ref{figure5})
with (\ref{figure4}) and seeing how the value of the constraint
mirrors the error in the perennial. It should be noted that the
proportionality factor between the value of the constraint and the
error is a function of the dynamical variables and therefore the value
of the constraint should only be taken as an indicator, not an exact
measure of the error. However, it is a good indicator when one carries
out convergence studies, since there the dynamical variables do not
change in value significantly from one resolution to the next and the
constraint diminishes as one converges to the continuum.

\begin{figure}[htbp]
  \centerline{\psfig{file=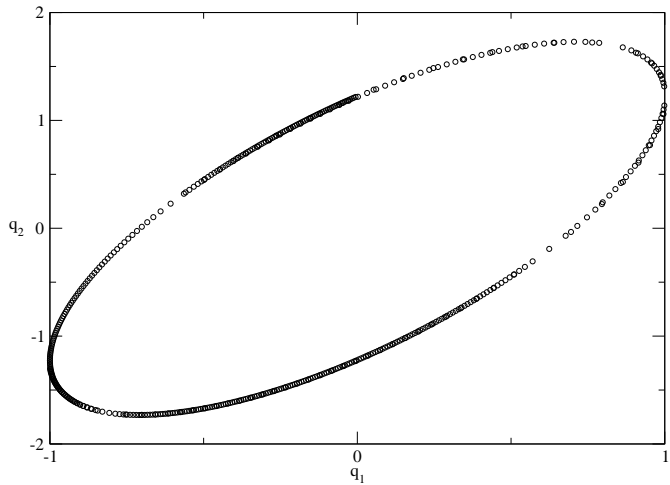,height=8.2cm}} \caption{The orbit
    in configuration space. As it is readily seen, the consistent
    discrete approach covers the entire available configuration space.
    This clearly exhibits that the approach is not a ``gauge fixing''.
    Gauge fixed approaches cannot cover the entire configuration space
    due to its compact nature. The dynamical changes in the value of
    the lapse can be seen implicitly through the density of points in
    the various regions of the trajectory. Also apparent is that the
    trajectory is traced on more than one occasion in various
    regions. Deviation from the continuum trajectory is not noticeable
    in the scales of the plot.}
  \label{figure3} \end{figure}

Figure (\ref{figure3}) shows the trajectory in configuration space.
As we see, the complete trajectory is covered by the discretized
approach.  This is important since many people tend to perceive the
consistent discretization approach as ``some sort of gauge fixing''.
This belief stems from the fact that when one gauge fixes a theory,
the multipliers get determined. In spite of this superficial analogy,
there are many things that are different from a gauge fixing. For
instance, as we discussed before, the number of degrees of freedom
changes.  In addition to this, this example demonstrates another
difference. If one indeed had gauge fixed this model, one would fail
to cover the entire available configuration space, given its compact
nature.

\begin{figure}[htbp]
  \centerline{\psfig{file=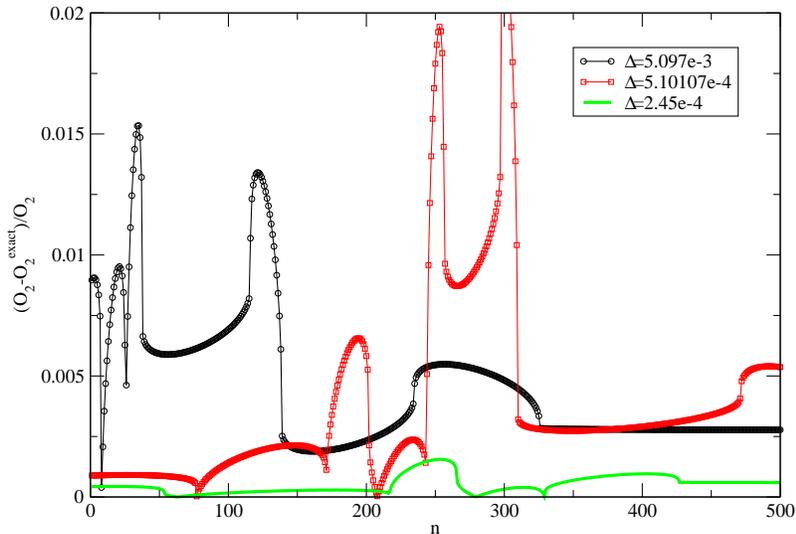,height=9.2cm}}
  \caption{Convergence of the method with increasing resolution. 
We display the relative error in one of the perennials of the theory
as a function of $n$. The
  two runs with the coarser resolutions are shown with one point out
  of every ten displayed. The finer run is shown with one point
  displayed out of every thirty. The range of $n$ displayed
  corresponds to a full trajectory along the ellipse in configuration
  space. So the improvement in accuracy is throughout the entire
  evolution with different levels of improvement at different points.
See the text as to why we fine tune the evolution steps for the
  various runs in the convergence study.}
  \label{figure6}
\end{figure}

An issue of interest in any numerical method is the concept of
convergence. Any reasonable numerical scheme should be such that one
has control on the approximation of the exact solution. Figure
(\ref{figure6}) shows the convergence in the error of the estimation
of the observable $O_2$. We see that one has convergence in the
traditional sense, i.e.  making the discretization step smaller lowers
the errors. However, one notes differences with the usual type of
convergence in the sense that here we have that it is not uniform as a
function of the evolution time.  One notes, for example, that at
isolated points some of the coarser runs have very low errors. To
understand this one needs to recall that in our approach discrete
expressions differ from the continuum ones as,
\begin{equation}
  O = O_{\rm continuum} +f(p,q) C
\end{equation}
with $C$ the constraints. It could happen that at a particular point
in a coarse evolution the constraint chances to take a value very
close to zero. Slightly finer resolution runs may not land on top of
that particular point and therefore will apparently have larger errors
in the region. Eventually, if one increases the resolution enough, one
will be able to achieve better accuracy than with the coarser run.
The reader may ponder why the parameter we have chosen to characterize
the discretization is listed with several digits of precision in the
convergence runs. The reason is the following: our algorithm exhibits
some sensitivity on the parameter in the following sense. If one
chooses slightly different values of $\Delta$, the way in which the
scheme will usually cover the phase space will be different.  For
instance, one choice may cover the configuration space ellipse in one
continuous sweep, another close choice may reverse course several
times before covering the ellipse. From a physical point of view this
is no problem, but from the point of view of comparing runs as a
function of $n$ one wishes to compare runs that behave in the same
way. Therefore when one ``halves the stepsize'' one may need to fine
tune by trial an error to make sure one is comparing evolutions that
behave similarly in terms of the (unphysical) parameter $n$.

\section{Applications in quantum cosmology}

To try to seek an application more connected with gravitational
physics, yet simple enough that we can solve things analytically, let
us consider a cosmological model. The model in question will be a
Friedmann model with a cosmological constant and a scalar field. To
make the solution of the model analytically tractable we will consider
a very massive scalar field (i.e. we will ignore the kinetic term of
the field in the action). We have actually solved this and other
models without this approximation numerically and obtained results
that are conceptually similar to the ones we present here.

The Lagrangian for the model, written in terms of Ashtekar's new
variables (there is no impediment in treating the model with
traditional variables if one wishes) is,
\begin{equation}
L= E\dot{A} + \pi \dot{\phi}- N E^2 (-A^2+(\Lambda +m^2 \phi^2)|E|)\label{lag}
\end{equation}
where $\Lambda$ is the cosmological constant, $m$ is the mass of the
scalar field $\phi$, $\pi$ is its canonically conjugate momentum and
$N$ is the lapse with density weight minus one. Here $E$ and $A$ are
the functions of time which are what remains of the triad and
connection for the homogeneous case. The appearance of $|E|$ in the
Lagrangian is due to the fact that the term cubic in $E$ is supposed
to represent the spatial volume and therefore should be positive
definite. In terms of the ordinary lapse $\alpha$ we have $\alpha = N
|E|^{3/2}$. The equations of motion and the only remaining
constraint (Hamiltonian) are
\begin{eqnarray}
\dot{A} &=& N (\Lambda+m^2\phi^2) {\rm sgn}(E) E^2\\
\dot{E} &=&2 N E^2 A\\
\dot{\phi}&=&0\\
\dot{\pi} &=&-2 N |E|^3 m^2 \phi\\
A^2&=&(\Lambda+m^2\phi^2) |E|\label{constr}
\end{eqnarray}

It immediately follows from the large mass approximation that
$\phi={\rm constant}$.  To solve for the rest of the variables, we
need to distinguish four cases, depending on the signs of $E$ and
$A$. Let us call $\epsilon={\rm sgn}(E)$ and $\chi={\rm sgn}(A)$. Then
the solution (with the choice of lapse $\alpha=1$) is,
\begin{eqnarray}
A&=&\chi \exp\left(\chi\epsilon t \sqrt{\Lambda+m^2\phi^2} \right)\\
E&=&\epsilon {\exp\left(2\chi\epsilon t \sqrt{\Lambda+m^2\phi^2}
\right) \over {\Lambda+m^2\phi^2} }
\end{eqnarray}

There are four possibilities according to the signs $\epsilon,\chi$.
If $\epsilon=\chi=1$ or $\epsilon=\chi=-1$ we have a universe that
expands.  If both have different signs, the universe contracts. This
just reflects that the Lagrangian is invariant if one changes the sign
of either $A$ or $E$ and the sign of time. It is also invariant if one
changes simultaneously the sign of both $A$ and $E$.

Let us turn to the observables of the theory (quantities that have
vanishing Poisson brackets with the constraint (\ref{constr}) and
therefore are constants of the motion). The theory has four phase
space degrees of freedom with one constraint, therefore there should
be two independent observables. Immediately one can construct an
observable $O_1=\phi$, since the latter is conserved due to the large
mass approximation. To construct the second observable we write the
equation for the trajectory,
\begin{equation}
{d\pi \over d A}= {-2 E m^2 \phi \over \Lambda+m^2 \phi^2} = -{2 A^2
m^2 \phi \over \left(\Lambda+m^2 \phi^2\right)^2} {\rm sgn} E
\end{equation}
where in the latter identity we have used the constraint. Integrating,
we get the observable,
\begin{equation}
O_2 = \pi + {2 \over 3} {m^2 \phi \over \left(\Lambda+m^2
\phi^2\right)^2} A^3 {\rm sgn} E \label{obse1}
\end{equation}
and using the constraint again we can rewrite it,
\begin{equation}
O_2 = \pi + {2 \over 3} {m^2 \phi \over \Lambda +m^2 \phi^2} A E.
\end{equation}

Although the last two expressions are equivalent, we will see that
upon discretization only one of them becomes an exact observable of
the discrete theory.

We consider the evolution parameter to be a discrete variable. Then the
Lagrangian becomes
\begin{equation}
L(n,n+1)=E_n (A_{n+1}-A_n)+ \pi_n (\phi_{n+1}-\phi_n) -N_n E_n^2
(-A^2_n+ (\Lambda+m^2 \phi^2_n) |E_n|)
\end{equation}

The discrete time evolution is generated by a canonical transformation
of type 1 whose generating function is given by $-L$, viewed as a
function of the configuration variables at instants $n$ and $n+1$. The
canonical transformation defines the canonically conjugate momenta to
all variables.  The transformation is such that the symplectic
structure so defined is preserved under evolution. The configuration
variables will be $(A_n,E_n,\pi_n,\phi_n,N_n)$ with canonical momenta
$(P^A_n,P^E_n,P^\phi_n,P^\pi_n,P^N_n)$ defined in the traditional
fashion by functional differentiation of the action with respect to
the canonical variables. We do not reproduce their explicit expression
here for reasons of brevity, the reader can consult them in reference
\cite{cosmo}. The definitions of the momenta can be combined in such a
way as to yield a simpler evolution system,
\begin{eqnarray}
A_{n+1}-A_n&=&N_n (P^A_{n+1})^2_n(\Lambda+m^2 \phi^2_n){\rm sgn}
P^A_{n+1},
\label{anm1}\\
P^A_{n+1}-P^A_n &=&2 N_n A_n  (P^A_{n+1})^2\label{pnm1}\\
\phi_{n+1}-\phi_n&=&0\label{phinm1}\\
P^\phi_{n+1}-P^\phi_n &=&-2 N_n (P^A_{n+1})^2 m^2 \phi_n |P^A_{n+1}|,\\
0&=&-A^2_n+ (\Lambda+m^2 \phi^2_n) |P^A_{n+1}|,\label{cons2}
\end{eqnarray}
and the phase space is now spanned by $A_n,P^A_n,\phi_n,P^\phi_n$. 

From (\ref{pnm1}) we determine,
\begin{equation}
P^A_{n+1}={1 +\xi \sqrt{1-8P^A_n A_n N_n}\over 4 A_n N_n},
\end{equation}
where $\xi=\pm 1$ and we will see the final solution is independent of
$\xi$.  Substituting in (\ref{cons2}) and solving for the lapse we
get,
\begin{equation}
N_n = {\left[-P^A_n \left(\Lambda+m^2 \phi_n^2\right)+A_n^2{\rm
sgn}P^A_n\right] \left(\Lambda+m^2\phi_n^2\right)\over 2
A_n^5}.\label{lapse}
\end{equation}

Let us summarize how the evolution scheme presented actually
operates. Let us assume that some initial data $A^{(0)},P^A_{(0)}$,
satisfying the constraints of the continuum theory, are to be
evolved. The recipe will consist of assigning $A_0=A^{(0)}$ and
$P^A_1=P^A_{(0)}$. Notice that this will automatically satisfy
(\ref{cons2}).  In order for the scheme to be complete we need to
specify $P^A_0$. This is a free parameter. Once it is specified, then
the evolution equations will determine all the variables of the
problem, including the lapse. Notice that if one chooses $P^A_0$ such
that, together with the value of $A_0$ they satisfy the constraint,
then the right hand side of the equation for the lapse (\ref{lapse})
would vanish and no evolution takes place. It is clear that one can
choose $P^A_0$ in such a way as to make the evolution step as small as
desired.

The equation for the lapse (\ref{lapse}) implies that the lapse is a
real number for any real initial data. But it does not immediately
imply that the lapse is positive. However, it can be shown that the
sign of the lapse, once it is determined by the initial configuration,
does not change under evolution. The proof is tedious since one has to
separately consider the various possible signs of $\epsilon$ and
$\chi$.This is an important result. In spite of the simplicity of the
model, there was no a priori reason to expect that the construction
would yield a real lapse.  Or that upon evolution the equation
determining the lapse could not become singular or imply a change in
the sign of the lapse, therefore not allowing a complete coverage of
the classical orbits in the discrete theory.

In general the discrete theory, having more degrees of freedom than
the continuum theory, will have more constants of the motion than 
observables in the continuum theory. In this example, the discrete
theory has four degrees of freedom. One can find four constants of the
motion. One of them we already discussed. The other one is $\phi$. 
The two other constants of the motion can in principle be worked
out. One of them is a measure of how well the discrete theory 
approximates the continuum theory and is only a function of the 
canonical variables (it does not depend explicitly on $n$). The
constant of the motion is associated with the canonical transformation
that performs the evolution in $n$. It is analogous to the Hamiltonian
of the discrete theory. The expression can be worked out as a power
series involving the discrete expression of the constraint of the 
continuum theory. This constant of motion therefore vanishes in
the continuum limit. The other constant of the motion also vanishes
in the continuum limit.

That is, we have two of the constants of the motion that reduce to the
observables of the continuum theory in the continuum limit and two
others that vanish in such limit. The discrete theory therefore
clearly has a remnant of the symmetries of the continuum theory. The
canonical transformations associated with the constants of the motion
which have non-vanishing continuum limit map dynamical trajectories to
other trajectories that can be viewed as different choices of lapse in
the discrete theory. This is the discrete analog of the
reparameterization invariance of the continuum theory. As we will see
soon the lapse in the discrete theory is determined up to two
constants. The choice of these two constants is the remnant of the
reparameterization invariance of the continuum theory that is present
in the discrete theory.

\begin{figure}
\centerline{\psfig{file=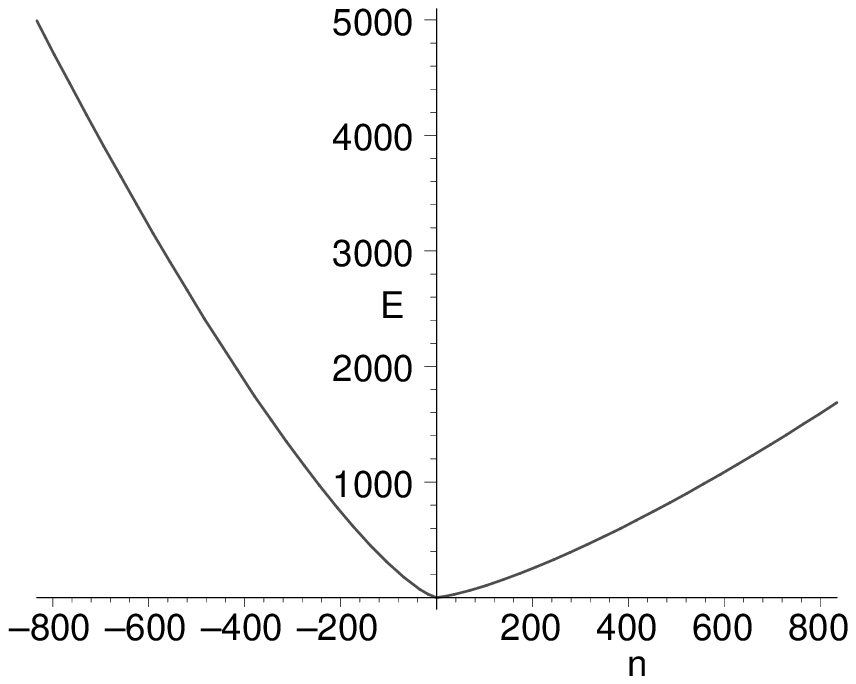,height=5cm},\psfig{file=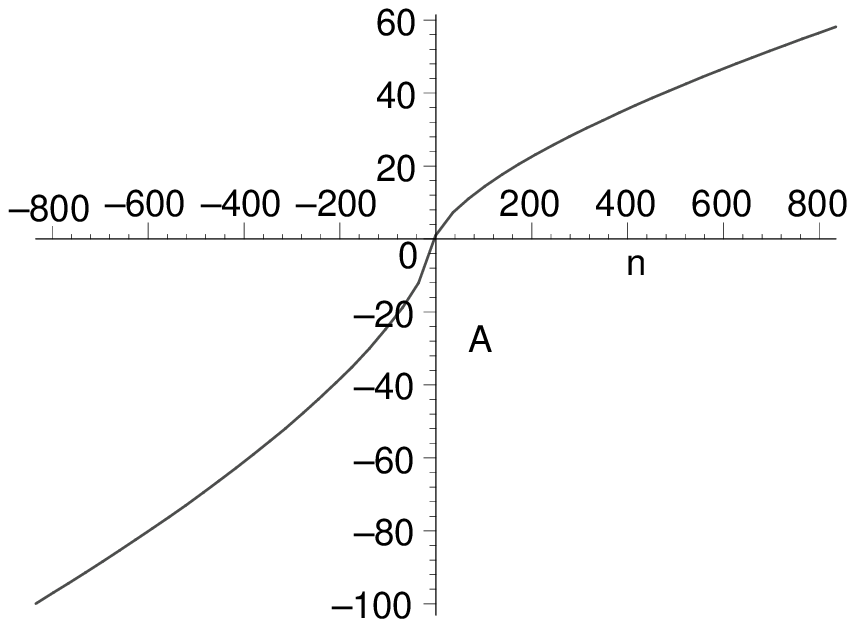,height=5cm}}
\caption{The discrete evolution of the triad $E$ and the connection $A$ as a
function of the discrete evolution parameter $n$. We have chosen initial 
conditions that produce a positive branch of $A$ for $n>0$ and a negative
branch for $n<0$. For the triad we chose both branches positive.}
\label{eadis}
\end{figure}

\begin{figure}
\centerline{\psfig{file=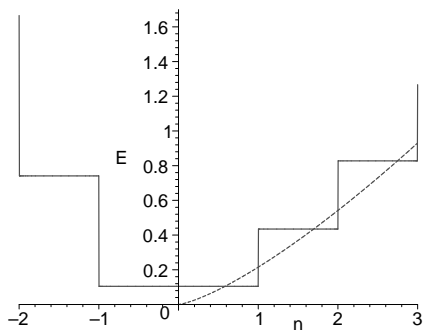,height=5cm}}
\caption{The approach to the singularity in the discrete and continuum 
solutions. The discrete theory has a small but non-vanishing triad at $n=0$
and the singularity is therefore avoided.}
\label{edism}
\end{figure}

Figure (\ref{eadis}) shows the comparison of the discrete evolution of
the model with the exact solution of the continuum theory.  As we see
the discrete theory approximates the continuum theory very well.
Figure (\ref{edism}) blows up the region of the evolution close
to the singularity. As it can be seen the discrete theory evolves
through the singularity. Emerging on the other side the evolution 
has a different value for the lapse and therefore a different
time-step. This could be used to implement the proposal of Smolin
that physical constants change when one tunnels through the 
singularity (in lattice theories the lattice spacing determines 
the ``dressed'' value of physical constants). See \cite{smolin}
for further discussion of this point.

To quantize the theory one has to implement the equations of 
evolution via a unitary transformation. Details of the derivation
can be found in \cite{cosmo}. The result is 
\begin{eqnarray}
<A_1,\phi_1,n||A_2,\phi_2,n+1>&=&
\delta(\phi_1-\phi_2) \exp\left(i{{\rm sgn}(A_1) A_2^2 (A_1-A_2)
\over \Theta}\right) \sqrt{{|A_1|\over \pi \Theta}}.
\end{eqnarray}

It should be noted that not all canonical transformations correspond
to unitary evolutions at the quantum level. If the canonical 
transformation defines an isomorphism between the phase spaces
at levels $n$ and $n+1$, then one can show that the canonical
transformation can be implemented by an isometry at a quantum
level. If the isomorphism is an automorphism then the canonical
transformation can be implemented as a unitary transformation.
A good discussion of canonical transformations in quantum
mechanics can be found in Anderson \cite{Anderson}.

With the unitary transformation introduced above one can answer any
question about evolution in the Heisenberg picture for the model in
question. One could also choose to work in the Schr\"odinger picture,
evolving the states. Notice that the wavefunctions admit a genuine
probabilistic interpretation. This is in distinct contrast with the
usual ``naive Schr\"odinger interpretation'' of quantum cosmology
which attempted to ascribe probabilistic value to the square of a
solution of the Wheeler--DeWitt equation (see \cite{Kuchar} for a
detailed discussion of the problems associated with the naive
interpretation).

An interesting aspect of this quantization is that for any
square-integrable wavefunction and for any value of the parameter $n$,
the expectation value of $(P^A_n)^2$, and therefore that of $E^2$ is
non-vanishing, and so are the metric and the volume of the
slice. Therefore quantum mechanically one never sees a singularity.
The mechanism for elimination of the singularity is quite distinct
from the one encountered in loop quantum cosmology \cite{Bojowald}.
The reader may ask how generic is our mechanism. Are singularities
always avoided? Singularities are avoided (simply because the a
lattice point generically will not coincide with them) provided that
they do not occur at a boundary of the phase space. If they occur at a
boundary, it is guaranteed that a lattice point will coincide with the
singularity and therefore it will not be avoided. In the example we
showed the singularity does not occur on the boundary.  However, it
appears this is a special feature of the large mass approximation in
the scalar field. For other models we have studied, at least worked
out in the Ashtekar variables, the connection diverges at the
singularity and therefore the singularity is at one boundary of the
phase space. This can be remedied by changing variables before
discretizing so that the singularity is not on a boundary, but it is
not what is the direct outcome of discretizing the theory in the
Ashtekar variables. An alternative that also eliminates the
singularity is to rewrite things in terms of holonomic variables as is
done in loop quantum cosmology.

As we will discuss in the next section, it is more desirable to
introduce a quantization that is relational in nature. This is because
the evolution parameter $n$ cannot be observed physically and
therefore is not a good choice of time to have a quantization that is
physical. Having implemented the evolution equations as a unitary
transformation and having a probabilistic interpretation for the
wavefunctions is all that is needed to work out a relational
quantization in detail without any of the conceptual problems
that were encountered in the past (see Page and Wootters \cite{PaWo}
and the critique of their work by Kucha\v{r} \cite{Kuchar}).

To define a time we therefore introduce the conditional probabilities,
``probability that a given variable have a certain value when the
variable chosen as time takes a given value''.  For instance, taking
$A$ as our time variable, let us work out first the probability that
the scalar field conjugate momentum be in the range $\Delta P^\phi=
[P^\phi_{(1)},P^\phi_{(2)}]$ and ``time'' is in the range $\Delta
A=[A_{(1)}, A_{(2)}]$ (the need to work with ranges is because we are
dealing with continuous variables). We go back to the naive
quantization and recall that the wavefunction $\Psi[A,\phi,n]$ in the
Schr\"odinger representation admits a probabilistic interpretation.
One can also define the amplitude $\Psi[A,P^\phi,n]$ by taking the
Fourier transform. Therefore the probability of simultaneous
measurement is,
\begin{equation}
P_{\rm sim}(\Delta P^\phi,\Delta A) = \lim_{N\to \infty} {1 \over N}
\sum_{n=0}^N \int_{P^\phi_{(1)},A_{(1)}}^{P^\phi_{(2)},A_{(2)}}
\Psi^2[A,P^\phi,n] dP^\phi dA. \label{psim}
\end{equation}
We have summed over $n$ since there is no information about the
``level'' of the discrete theory at which the measurement is
performed, since $n$ is just a parameter with no physical
meaning. With the normalizations chosen if the integral in $P^\phi$
and $A$ were in the range $(-\infty,\infty)$, $P_{\rm sim}$ would be
equal to one.

To get the conditional probability $P_{\rm cond}(\Delta P^\phi|\Delta
A)$, that is, the probability that having observed $A$ in $\Delta A$
we also observe $P^\phi$ in $\Delta P^\phi$, we use the standard
probabilistic identity
\begin{equation}
P_{\rm sim}(\Delta P^\phi,\Delta A) =
P(\Delta A) P_{\rm cond}(\Delta P^\phi|\Delta A)
\end{equation}
where $P(\Delta A)$ is obtained from expression (\ref{psim})
taking the integral on $P^\phi$ from $(-\infty,\infty)$. We therefore
get 
\begin{equation}
P_{\rm cond}(\Delta P^\phi|\Delta A)=
{\lim_{N\to \infty} {1 \over N}
\sum_{n=0}^N \int_{P^\phi_{(1)},A_{(1)}}^{P^\phi_{(2)},A_{(2)}}
\Psi^2[A,P^\phi,n] dP^\phi dA \over
\lim_{N\to \infty} {1 \over N}
\sum_{n=0}^N \int_{-\infty,A_{(1)}}^{\infty,A_{(2)}}
\Psi^2[A,P^\phi,n] dP^\phi dA}.  
\end{equation}
Notice that all the integrals are well defined and the resulting 
quantity behaves as a probability in the sense that integrating 
from $(-\infty,\infty)$ in $P^\phi$ one gets unity.

Introducing probabilities is not enough to claim to have completed
a quantization. One needs to be able to specify what happens to the
state of the system as a measurement takes place. The most natural
reduction postulate is that,
\begin{equation}
|\psi> \rightarrow {\Pi_{P^\phi,A} |\phi> \over
\sqrt{|<\phi|\Pi_{P^\phi,A}|\phi>|}},
\end{equation}
where 
\begin{equation}
\Pi_{P^\phi,A} = \sum_{n=0}^N |P^\phi,A,n><P^\phi,A,n|.
\end{equation}

The model considered is too simple to test too much of the framework,
however, we have shown that one can work out in detail the 
discrete treatment at a classical an quantum mechanical level without
conceptual obstructions. It is a big leap to claim that because
everything worked well for such a simple model these ideas will
succeed in full GR. Currently we are working in detail the
discretization of Gowdy cosmologies, which have field theoretic
degrees of freedom. Recently achieved success \cite{Gowdy} in such 
models greatly enhances our confidence in the ability of this
scheme to discretize general relativity.

For now, we would like to take a glimpse at some possibilities
that the framework will introduce in the full theory, we do so in
the next section.

\section{Applications in quantum gravity}

Having made the case that one can approximate general relativity by a
discrete theory that is constraint-free allows us to make progress in
many aspects of quantum gravity. Most of the hard conceptual problems
that one faces in canonical quantum gravity are related to the
presence of constraints in the theory. For an unconstrained theory,
most obstructions are eliminated. One of the first obstructions we can
deal with is the ``problem of time''. To a certain extent we have
shown that this is possible in Rovelli's example (which has a
``problem of time'') and in the cosmological model. But actually
progress is possible in a more generic sense. One can, for instance,
implement the relational time that was proposed by Page and Wootters
\cite{PaWo} in the full theory. The idea consists in quantizing the
theory by promoting all variables to quantum operators, unlike the
usual Schr\"odinger quantization in which a variable called ``time''
is kept classical. One then picks from among the quantum operators one
that we will call ``clock''. It should be emphasized that it is a
quantum operator and therefore will have an expectation value,
fluctuations, etc. One then computes conditional probabilities for the
other variables in the theory to take certain values when the
``clock'' variable has a given value.

The resulting conditional probabilities do not evolve according to a
Schr\"odinger equation. If one chooses as ``clock'' a variable that
behaves close to classicality (small fluctuations) then one can show
that the conditional probabilities evolve approximately by a
Schr\"odinger evolution. But as the fluctuations in the clock variable
increase there appear corrective terms in the evolution equations. The
fist kind of corrective terms were evaluated in \cite{GaPoPunjp} and
have the form of a Lindblad type of evolution. A feature of the
evolution implied by the use of ``real clocks'' as we are considering
is that it is not unitary. In general pure states evolve into mixed
states. This is easy to understand in our discrete approach. There we
saw that evolution in terms of the discrete parameter $n$ was
implemented by a canonical transformation. Upon quantization evolution
is implemented by a unitary operator. However, the real clock variable
we choose will in general have a probability distribution with a
certain spread in $n$.  If evolution is unitary in $n$ then it cannot
be unitary in the clock variable since in general for a given value of
the clock variable one will have a superposition of states with
different values of $n$.

This lack of unitarity could in principle be experimentally
observable.  We have estimated its magnitude in \cite{GaPoPuprl}. In
order to make an estimate one needs to make a model of what is the
``most classical'' clock one can construct. To do this we borrowed
from ideas of Ng and Van Dam, Amelino-Camelia and more recently Lloyd
and collaborators\cite{ngac}. They start with the observation of
Salecker and Wigner \cite{SaWi} that the accuracy of a clock is 
limited by its mass $\delta T > \sqrt{T/M}$ (in units of $\hbar=c=1$)
where $T$ is the time to be measured. Then they argue that the
maximum amount of mass one can concentrate can be achieved in 
a black hole. If the clock is a black hole its accuracy is given
by its quasinormal frequencies, $\delta T > T_{P}^2 M$. where 
$T_P$ is Planck's time. Combining the two inequalities we get, 
$\delta T \sim T_P \sqrt[3]{T/T_P}$

If one now considers a model system consisting of two quantum
mechanical levels and estimates the lack of coherence induced by the
use of the relational time one finds that it that the elements of the
density matrix that are off-diagonal in the energy basis decay
exponentially. The exponent is given by $t_P^{(4/3)} T^{(2/3)}
\omega_{12}^2$ where $T$ is the time we observe the system and
$\omega_{12}$ is the Bohr frequency between both levels. Given the
presence of the $t_P$ factor, this effect is unobservable for almost
all experimental situations. An exception could be ``macroscopic''
quantum states, like the ones that are starting to become available
through Bose-Einstein condensates. Current technology does not produce
states that are ``macroscopic enough'' for the effect to be
observable, and it remains to be seen if future technologies could
produce such states (and keep them free of ambiental decoherence
effects) for this effect to be observable (for an independent
discussion see \cite{SiJa}.

One place where the fundamental decoherence could play a role
is in the black hole information puzzle. Black holes take a
very long time to evaporate and therefore there is a chance
for the decoherence to build up. The question is: does it build
up enough to wipe out all coherence from the states before
the black hole would have done the same via Hawking evaporation?
We recently estimated the effect by considering a very naive
model of a black hole consisting of two energy levels 
separated by $k T$ with k the Boltzmann constant and $T$ the
temperature predicted by Hawking for the black hole. We found
out that \cite{piombino}
\begin{equation}
|\rho_{12}(T_{\rm max})| \sim |\rho_{12}(0)| \left({M_{\rm Planck}
 \over M_{\rm BH}}\right)^{2 \over 3}
\end{equation}
where $\rho_{12}$ is an off-diagonal element of the density
matrix of the state of interest at the time the black hole would
have evaporated. For a Solar sized black hole
the magnitude of the off diagonal element is $10^{-28}$, that is,
it would have de facto become a mixed state even before invoking
the Hawking effect. The information paradox is therefore 
unobservable in practice. It should be noted that this estimate
is an optimistic one. In reality clocks fare much worse than
the estimate we worked out and the fundamental decoherence will
operate even faster than what we consider here.

\section{Connections with continuum loop quantum gravity}

In spite of the possibilities raised by the discrete approach, some
readers may feel that it forces us to give up too much from the
outset. This was best perhaps captured by Thomas Thiemann
\cite{Thiemann}, who said ``While being a fascinating possibility,
such a procedure would be a rather drastic step in the sense that it
would render most results of LQG obtained so far obsolete''. Indeed,
the kinematical structure built in loop quantum gravity, with a
rigorous integration measure and the natural basis of spin foam states
appear as very attractive tools to build theories of quantum
gravity. We would like to discuss how to recover these structures in
our discrete approach.

To make contact with the traditional kinematics of loop quantum
gravity, we consider general relativity and discretize time but
keep space continuous, and we proceed as in the consistent 
discretization approach, that is, discretizing the action and 
working out the equations of motion. We start by considering the
action written in terms of Ashtekar variables
\cite{asle},
\begin{equation}
  S=\int dt d^3x \left(\tilde{P}^a_i F_{0a}^i -N^a C_a -N C\right) 
\end{equation}
where $N,N^a$ are Lagrange multipliers, $\tilde{P}^a_i$ are densitized
triads, and the diffeomorphism and
Hamiltonian constraints are given by,
$
  C^a=\tilde{P}^a_i F_{ab}^i$, 
$  C ={\tilde{P}^a_i \tilde{P}^b_j \over \sqrt{\rm det} q}
\left(\epsilon^{ijk} F_{ab}^i -
(1+\beta^2) K_{[a}^i K_{b]}^j \right)$
where $\beta K_a^i\equiv \Gamma_a^i-A_a^i $ and $\Gamma_a^i$ is the
spin connection compatible with the triad, and $q$ is the three
metric. We now proceed to discretize time.  The action now reads,
\begin{eqnarray}
  S&=&\int dt d^3x \left[{\rm Tr}\left( \tilde{P}^a
\left(A_a(x) -V(x) A_{n+1,a}(x) V^{-1}(x)+\partial_a(V(x)) V^{-1}(x)\right)
\right)\right.\\
&&\left.-N^a C_a -N C+ \mu \sqrt{\rm det} q  {\rm Tr}\left(V(x)V^\dagger(x)-1\right)\right] \nonumber
\end{eqnarray}
In the above expression $V(x)=V_I T^I$ is the parallel transport
matrix along a time-like direction and $F_{0a}$ is approximated by the
holonomy along a plaquette that is finite in the ``time-like''
direction and infinitesimal in the ``space-like'' direction and
$T^0=1/\sqrt{2}$ and $T^a=-i\sigma^a/\sqrt{2},a=1..3$ and $\sigma$'s
are the Pauli matrices and the coefficients $V_I$ are real. We have
omitted the subscript $n$ to simplify the notation and kept it in the
quantities that are evaluated at $n+1$. The last term involves a
Lagrange multiplier $\mu$ and is present in order to enforce the fact
that the parallel transport matrices are unitary.  We notice that the
$SU(2)$ gauge invariance is preserved in the semi-discrete theory.
This in turn implies that Gauss' law for the momentum canonically
conjugate to the connection, $\tilde{E}^a_{n+1} \equiv V^{-1}
\tilde{P}^a V$ is preserved automatically upon evolution. Remarkably,
although the theory does not have a diffeomorphism constraint, one can
show that the quantity that in the continuum would correspond to the
diffeomorphism constraint is a conserved quantity (to intuitively see
this, notice that the action is invariant under (time independent)
spatial diffeomorphisms and $n$ and $n+1$). One can then impose that
this quantity vanish and this requirement is consistent with
evolution. One would therefore have a theory with the same constraints
as kinematical loop quantum gravity and with an explicit evolution
scheme as is usual in consistently discretized theories.

In reference \cite{difeo} we have actually worked out the procedure
in detail for the case of $2+1$ dimensions, treating gravity as a 
BF theory. The procedure reproduces the physical space for the
theory of traditional quantizations. 

Summarizing, one can apply the consistent discretization approach
by discretizing only time. The resulting theory has no constraints,
but the diffeomorphism constraint can be introduced additionally,
so the resulting theory has the same kinematics as loop quantum
gravity. The dynamics is implemented via a unitary operator.
As in the full discrete approach, one can envision bypassing many
of the hard conceptual questions of canonical quantum gravity,
for instance introducing a relational notion of time. This can
actually be seen as a concrete framework to implement loop
quantum gravity numerically, since it is not expected that one
will be able to work out things analytically in the full case.

\section{Conclusions}

The consistent discretization approach is emerging as an
attractive technique to handle discrete general relativity,
both at a classical and at a quantum mechanical level. In 
quantum gravity, since it does away with the constraints it
solves in one sweep some of the hardest conceptual issues of
canonical quantization. In classical general relativity there
is a growing body of evidence that suggests that the 
discretizations work numerically and approximate the
continuum theory (including its constraints) in a convergent
and stable way. Having discrete theories with these properties
classically is a desirable point of view for any effort 
towards quantization. The discrete approach also yields 
directly computable evolutions quantum mechanically opening
the possibility for concrete numerical quantum gravity
calculations.

\section{Acknowledgements}
This work was supported in part by grants NSF-PHY0244335,
and by funds of the Horace C. Hearne Jr. Institute for Theoretical
Physics and CCT-LSU.


\begin{references}
\bibitem{Marsd} A. Lew, J. Marsden, M. Ortiz, M. West, in
``Finite element methods: 1970's and beyond'', L. Franca, T. 
Tezduyar, A. Masud, editors, CIMNE, Barcelona (2004).
\bibitem{mclc} R. McLachlan, M. Perlmutter 
 ``Integrators for nonholonomic mechanical systems'' preprint (2005).
\bibitem{DiGaPu} C. Di Bartolo, R. Gambini, J. Pullin,
Class. Quan. Grav. 19, 5275 (2002).
\bibitem{GaPu} R. Gambini and J. Pullin, Phys. Rev. Lett. 90, 021301,
(2003).
\bibitem{jmp}
  C.~Di Bartolo, R.~Gambini, R.~Porto and J.~Pullin,
  J.\ Math.\ Phys.\  {\bf 46}, 012901 (2005)
  [arXiv:gr-qc/0405131].
\bibitem{Rovelli}
C.~Rovelli,
Phys.\ Rev.\ D {\bf 42}, 2638 (1990).
\bibitem{cosmo}
  R.~Gambini and J.~Pullin,
  Class.\ Quant.\ Grav.\  {\bf 20}, 3341 (2003)
  [arXiv:gr-qc/0212033].
\bibitem{smolin} 
R.~Gambini and J.~Pullin,
  Int.\ J.\ Mod.\ Phys.\ D {\bf 12}, 1775 (2003)
  [arXiv:gr-qc/0306095].
\bibitem{Anderson}A. Anderson, 
Annals Phys. {\bf 232},292-331 (1994) [arXiv:hep-th/9305054]. 
\bibitem{Kuchar}
K. Kucha\v{r}, ``Time and interpretations of quantum gravity'',
in ``Proceedings of the 4th Canadian conference on general
relativity and relativistic astrophysics'', G. Kunstatter, D.
Vincent, J. Williams (editors), World Scientific, Singapore
(1992). Available online at 
http://www.phys.lsu.edu/faculty/pullin/kvk.pdf
\bibitem{Bojowald}
  M.~Bojowald,
  Pramana {\bf 63}, 765 (2004)
\bibitem{PaWo} D.~N.~Page and W.~K.~Wootters,
Phys.\ Rev.\ D {\bf 27}, 2885 (1983); W. Wootters, Int. J. Theor. Phys.
{\bf 23}, 701 (1984); D. N. Page, ``Clock time and entropy'' in
``Physical origins of time asymmetry'', J. Halliwell, J. Perez-Mercader,
W. Zurek (editors), Cambridge University Press, Cambridge UK, (1992).
\bibitem{Gowdy} R. Gambini, M. Ponce, J. Pullin, ``Consistent
discretizations of Gowdy cosmologies'' (in preparation).
\bibitem{GaPoPunjp}
  R.~Gambini, R.~Porto and J.~Pullin,
  New J.\ Phys.\  {\bf 6}, 45 (2004)
  [arXiv:gr-qc/0402118].
\bibitem{GaPoPuprl}
R.~Gambini, R.~A.~Porto and J.~Pullin,
  Phys.\ Rev.\ Lett.\  {\bf 93}, 240401 (2004)
  [arXiv:hep-th/0406260].
\bibitem{ngac}
G.~Amelino-Camelia,
Mod.\ Phys.\ Lett.\ A {\bf 9}, 3415 (1994)
[arXiv:gr-qc/9603014];
Y.~J.~Ng and H.~van Dam,
Annals N.\ Y.\ Acad.\ Sci.\  {\bf 755}, 579 (1995) 
[arXiv:hep-th/9406110];
Mod.\ Phys.\ Lett.\ A {\bf 9}, 335 (1994);
V. Giovanetti, S. Lloyd, L. Maccone, Science {\bf 306}, 1330 (2004);
S. Lloyd, Y. J. Ng, Scientific American, November 2004.
\bibitem{SaWi}
E. Wigner, Rev. Mod. Phys. {\bf 29}, 255 (1957);
H. Salecker, E. Wigner, Phys. Rev. {\bf 109}, 571 (1958).
\bibitem{SiJa}
 C.~Simon and D.~Jaksch,
Phys. Rev. {\bf A70}, 052104 (2004)
  [arXiv:quant-ph/0406007].
\bibitem{piombino}
R.~A.~Gambini, R.~Porto and J.~Pullin,
``Fundamental decoherence in quantum gravity,'', to 
appear in the proceedings of 2nd International Workshop DICE2004: From
Decoherence and Emergent Classicality to Emergent Quantum Mechanics,
Castello di Piombino, Tuscany, Italy, 1-4 Sep 2004, H. T. Elze, editor  [arXiv:gr-qc/0501027].
\bibitem{Thiemann} T.~Thiemann,
``The Phoenix project: Master constraint programme for loop quantum
gravity,''  arXiv:gr-qc/0305080.
\bibitem{asle}
A. Ashtekar, J. Lewandowski
Class.\ Quant.\ Grav.\  {\bf 21}, R53 (2004)
[arXiv:gr-qc/0404018].
\bibitem{difeo}
R.~Gambini and J.~Pullin,
  Phys.\ Rev.\ Lett.\  {\bf 94}, 101302 (2005)
  [arXiv:gr-qc/0409057].
\end{references}
\end{document}